\documentclass[twocolumn, secnumarabic, amssymb, nobibnotes, prapplied, superscriptaddress]{revtex4-2}

\newcommand{\ket}[1]{\left|#1\right\rangle}
\newcommand{\murm}{%
  \ifmmode
    \mathchoice
        {\hbox{\normalsize\textmu}}
        {\hbox{\normalsize\textmu}}
        {\hbox{\scriptsize\textmu}}
        {\hbox{\tiny\textmu}}%
  \else
   ~\textmu
  \fi
}

\usepackage{bbold}

\usepackage{lipsum}
\usepackage{amsfonts,amssymb,amsmath}
\usepackage{graphicx}
\usepackage{bm}
\usepackage{calc}
\usepackage[x11names]{xcolor}
\usepackage{siunitx}
\usepackage{tikz}
\usepackage{wasysym}
\usepackage{dsfont}
\usepackage{hyperref}
\usepackage{diagbox}
\usepackage{booktabs}
\usepackage{float}

\newcommand{\fmarki}{*}
\newcommand{\fmarkii}{\ensuremath{\dagger}}
\newcommand{\fmarkiii}{\ensuremath{\ddagger}}

\def\@fnsymbol#1{{\ifcase#1\or \fmarki\or \fmarkii\or \fmarkiii\or \fmarkiv\or \fmarkv\or \fmarkvi\or \fmarkvii\or \fmarkviii\or \fmarkix \else\@ctrerr\fi}}
\makeatother

\renewcommand{\fmarki}{b$_1$}
\renewcommand{\fmarkii}{b$_2$}
\renewcommand{\fmarkiii}{c$_3$}

\definecolor{color_comment}{rgb}{0.8, 0.3, 0.3}

\definecolor{color_out}{rgb}{0.7, 0.7, 0.7}

\definecolor{color_new}{rgb}{0.3, 0.8, 0.3}

\DeclareMathOperator{\sinc}{sinc}

\definecolor{color_comment}{rgb}{0.8, 0.3, 0.3}

\def\be{\begin{equation}}
\def\ee{\end{equation}}
\def\bea{\begin{eqnarray}}
\def\eea{\end{eqnarray}}

\def\ket#1{\mbox{$\left|#1\right\rangle$}}

\usepackage{standalone}
\graphicspath{{figures/}}

\begin{document}

\title{Optimization and characterization of laser excitation for quantum sensing  with single nitrogen-vacancy centres}

\author{Alejandro Martínez-Méndez}
\thanks{Both authors contributed equally to this work}
\author{Jesús Moreno-Meseguer}
\thanks{Both authors contributed equally to this work}
\affiliation{Departamento de Física - CIOyN, Universidad de Murcia, Murcia E-30071, Spain}

\author{Mariusz Mr\'{o}zek}
\affiliation{Marian Smoluchowski Institute of Physics, Jagiellonian University, Lojasiewicza 11, 30-348 Krakow, Poland}

\author{Adam Wojciechowski}%
\affiliation{Marian Smoluchowski Institute of Physics, Jagiellonian University, Lojasiewicza 11, 30-348 Krakow, Poland}
\author{Priya Balasubramanian}
\affiliation{Institute for Quantum Optics, Ulm University, Albert-Einstein-Allee 11, 89081 Ulm, Germany}
\affiliation{Center for Integrated Quantum Science and Technology (IQST), 89081 Ulm, Germany}
\author{Fedor Jelezko}
\email{Corresponding author: fedor.jelezko(at)uni-ulm.de}
\affiliation{Institute for Quantum Optics, Ulm University, Albert-Einstein-Allee 11, 89081 Ulm, Germany}
\affiliation{Center for Integrated Quantum Science and Technology (IQST), 89081 Ulm, Germany}

\author{Javier Prior}
\email{Corresponding author: javier.prior(at)um.es}
\affiliation{Departamento de Física - CIOyN, Universidad de Murcia, Murcia E-30071, Spain}

\begin{abstract}

In this work we present a comprehensive method of characterization and optimization of laser irradiation within a confocal microscope tailored to quantum sensing experiments using nitrogen-vacancy (NV) centres. While confocal microscopy is well-suited for such experiments, precise control and understanding of several optical parameters are essential for reliable single-emitter studies. 
We investigate the laser beam intensity profile, single-photon emission statistics, fluorescence response under varying polarization and saturation conditions, spectral characteristics, and the temporal profiles of readout and reinitialization pulses. The beam quality is assessed using the beam propagation factor $M^2$, determined via the razorblade technique. Optical fluorescence spectrum is recorded to confirm NV centre emission. To confirm single-emitter operation, we measure second-order autocorrelation function $g^{(2)}(\tau)$. Saturation behaviour is analysed by varying laser power and recording the corresponding fluorescence, while polarization dependence is studied using a half-wave ($\lambda/2$) plate. Temporal laser pulse profile is examined by modulating the power of an acousto-optic modulator. After optimizing all relevant parameters, we demonstrate the microscope’s capabilities in driving spin transitions of a single NV centre. This work establishes a straightforward and effective protocol for laser excitation optimization, enhancing the performance and reliability of NV-based quantum sensors.

\end{abstract}

\maketitle

\section{Introduction}

\begin{figure}[h]
\includegraphics[width=\linewidth]{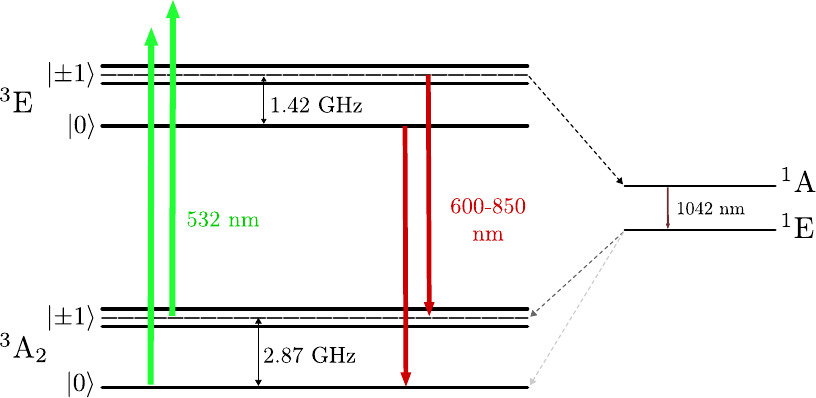}
\caption{Energy scheme of the NV centre. Transitions from the ground state $^3$A$_2$ to the excited state $^3$E can be induced by green laser excitation, with a spin conserving decay emitting red fluorescence. When exciting to the $\ket{\pm 1}$ state, there is a 30\% chance of decay through the intermediate singlets, where decay occurs on the infrared and can be filtered out for sensing experiments.}
\label{fig:energies}
\end{figure}

Negatively charged nitrogen-vacancy (NV) colour centres in diamond have established themselves at the forefront of solid-state spins for quantum sensing applications \cite{Glenn2018, GurudevDutt2007, Childress2006}. Their remarkable coherence times and functionality at room temperature stand out among other colour defects \cite{Gottscholl2021}. In particular, they are very simple to initialize and readout by means of green laser excitation and red fluorescence respectively, as shown in Fig. \ref{fig:energies}. NV centres act as spin triplets where the $\ket{\pm 1}$ spin states are degenerated and separated from the $\ket{0}$ state by a Zero-Field Splitting (ZFS) of 2.87 GHz, accessible via microwave (MW) irradiation. After excitation at 532 nm, relaxation can be spin-conserving directly to the ground state, or non-spin-conserving via a pair of intermediate-state singlets. The second path is favourable to the $\ket{\pm 1}$ state and decays to the $\ket{0}$ ground state. Continuous irradiation leads to a population pump into the $\ket{0}$ spin state, as the non-spin-conserving transition only favours this exchange. Comparison of fluorescence at the beginning and at the end of a measurement enables spin readout, with $\ket{\pm 1}$ emission reaching a top value of 30\% less fluorescent than infrared radiation coming from $\ket{0}$ excitation. This feature can be used to characterize the NV transitions with an Optically Detected Magnetic Resonance (ODMR) experiment \cite{Jarmola2016}.

NV centres are multimodal sensors that can be engineered in varying concentrations and configurations depending on their application. Low NV concentrations, below 0.1 part per billion (ppb) carbon atoms, embedded in highly pure diamond, can be used to detect single spins. However, high concentration ensembles, reaching part-per-million (ppm)-level NV configurations, can be created in bulk and nanodiamonds (NDs) for higher sensitivity \cite{Hong2013, Barry2024}. Each system will have its specific requirements, and therefore the microscopy techniques used for imaging and readout will change. Low concentrations require high optical resolution to be able to address individual centres, whereas good signal quality is prioritized for ensembles where spin contrast is reduced. These conditions match the main characteristics of confocal microscopy. Here, not only the excitation light is focused onto the sample, but also the emission will pass through a pinhole for spatial filtering. This results in a great increase in axial resolution together with a slight advantage in radial resolution compared to wide-field microscopy. The typical volume addressed in a modern confocal microscope is on the order of 0.1~$\mu\textrm{m}^3$ and, combined with a low NV concentration diamond sample, allows one to address individual NV centres. The improvement in resolution helps extract the fluorescence containing the spin information of the system without mixing the signal with external noise, usually in the form of background light, including fluorescence of other NV centres, which is quantified by the optical signal-to-noise ratio (SNR). SNR is of the utmost importance when performing quantum sensing experiments with NV centres since it will determine the quality of the signal received. Scanning of the sample in a confocal microscope, typically using piezoelectric nanopositioners, will produce a non-invasive 3D reconstruction \cite{Garsi2024}, making it ideal for mapping of NV concentration across different samples. 

Despite this natural accessibility, confocal microscopes must be complemented with additional features to be deemed optimal for quantum sensing experiments with NV centres. Most notably, green laser excitation needs to fulfil several requirements to extract the maximum performance from an experiment. A pulsed character for the laser is mandatory since the beam will be used not only for imaging but also for spin readout and coherent control through pulsed protocols \cite{Bhave2015}. Additionally, reduced focal spots will ensure the correct initialization and readout of single emitters \cite{Shields2015}, and polarization control \cite{Alegre2007, Hanson2006} will help extract the maximum fluorescence from the optical dipoles of the NV centre. Finally, the amount of laser power needs to be characterized for a sample in order to avoid saturation of the colour centres. 

In section II of this paper, we introduce theoretically the characteristics of Gaussian beams and the measurement protocol of the intensity profile, the polarization-dependent fluorescence and saturation behaviour of NV centres and the working principle of Acousto-Optic modulators. The confocal microscope employed for this work is presented in section III. The results of our protocol of measurements are shown in section IV. Finally, section V discusses the capabilities of the microscope for quantum sensing experiments after the protocol is applied. 

\section{Theoretical frame}

\subsection{Gaussian beams}

In confocal microscopy, the use of a Gaussian-profile laser beam is essential to achieve high resolution imaging and precise optical sectioning. This is because Gaussian beams can be focused on a minimally sized spot more effectively than other light distributions, thereby enhancing lateral resolution and allowing for accurate depth discrimination (axial resolution) and improved contrast. As discussed in the following, Gaussian beams are particularly advantageous because their intensity distribution follows a predictable bell-shaped curve, leading to efficient focusing and optimal interaction with the sample.

\begin{figure}[t]
\includegraphics[width=\linewidth]{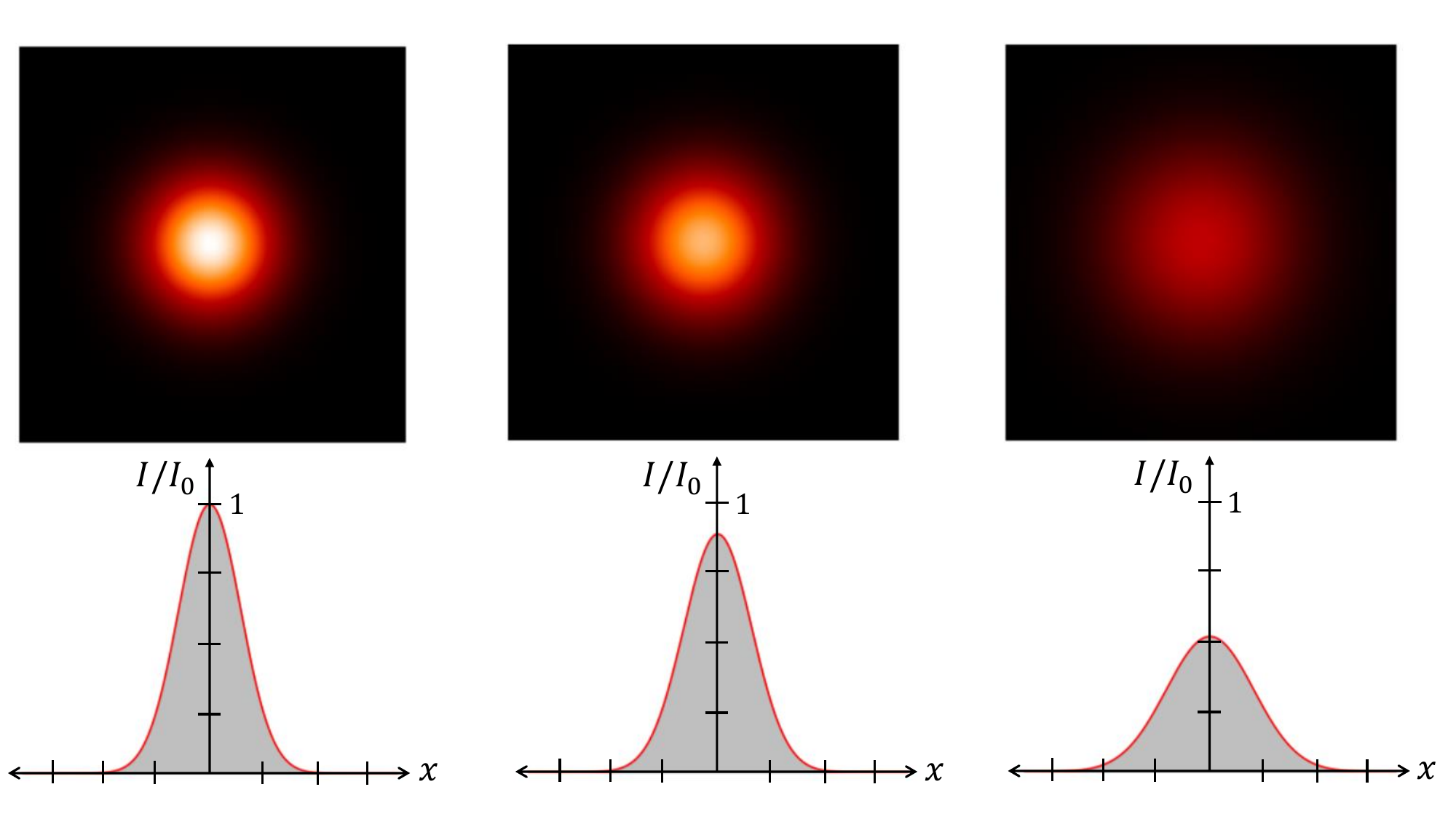}
\caption{Normalized Gaussian beam intensity $I$/$I_0$ as a function of the radial distance $\rho$ at different axial distances. From left to right: $z$ = 0, $z$ = $2z_0$, $z$ = $5z_0$. The amplitude decays and the signal gets broader as it distances from the focal plane.}
\label{fig:beam1}
\end{figure}

Gaussian beams, also known as TEM$_{00}$, are the simplest propagation mode of the light. The equations governing its distribution can be obtained by solving the Helmholtz Equation under the paraxial approximation. The intensity of this field propagating in the z axis is given by
\begin{equation}
I (\rho,z) = I_0 \left[ \frac{W_0}{W(z)}\right]^2 e^{-\frac{2 \rho^2}{W^2(z)}}.
\label{eq:int}
\end{equation}

\begin{figure}[h]
    \includegraphics[width=\linewidth]{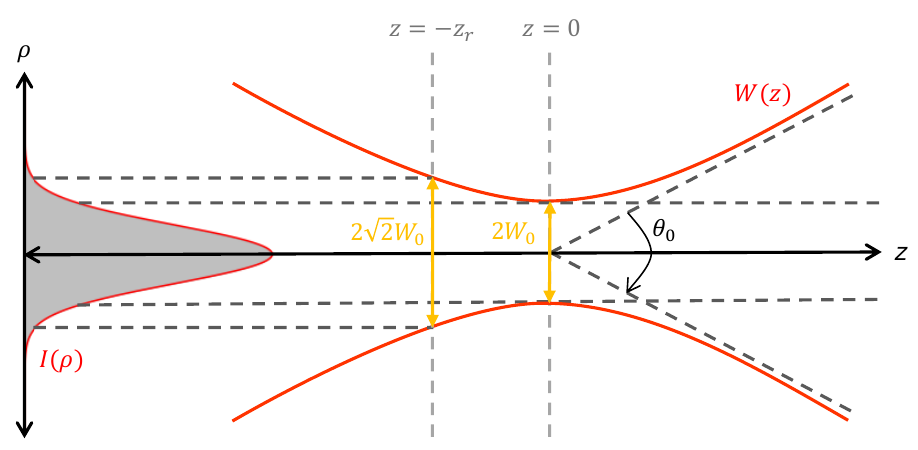}
    \centering
    \caption{Beam width $W(z)$ as function of distance z (Orange lines). $W_0$ describes the minimum radius of the beam and $\theta_0$ the convergence angle. The Gaussian intensity profile is depicted on the left, and it defines the position of $z_R$.}
    \label{fig:Propagation}
\end{figure}

Approximately 86\% of the total power is carried inside a circular cross section of radius $\rho(z) = W(z)$. In other words, intensity in the transversal plane drops by a factor of 1/$e^2$ at a distance $\rho(z) = W(z)$ from the z axis. This makes it very convenient to refer to W(z) as the beam radius or the beam width, 
\begin{equation}
W(z)  = W_0 \sqrt{1+\left( \frac{z}{z_R}\right)^2},
\label{eq:beam}
\end{equation}
where $z_R$ is the so-called Rayleigh parameter, and is related to the beam geometry as
\begin{equation}
W_0 = \sqrt{\frac{\lambda z_R}{\pi}}.
\label{eq:w0}
\end{equation}
$W_0$ is called the waist radius, and $2W_0$ is the spot size. The beam size is roughly kept inside the Rayleigh range $[-z_R, z_R]$, increasing to a value of $W( \pm z_R) = \sqrt2 \ W_0$,  before starting to increase linearly, as 

\begin{equation}
\frac{d }{dz}  W(z)= \frac{W_0 }{z_R}  =  \theta_0,
\label{eq:div}
\end{equation}
where $\theta_0$ is the divergence of the beam, as shown in \ref{fig:Propagation}. Using Eqs. \ref{eq:w0} and \ref{eq:div}:

\begin{equation}
\frac{W_0 \theta_0}{\lambda/\pi} = \frac{\pi W_0^2}{\lambda z_R} = 1.
\label{eq:m1}
\end{equation}

For an ideal Gaussian beam, the product of the waist radius $W_0$ and the angular divergence $\theta_0$ has a fixed, minimum value.

\subsubsection{Beam quality factor}

Gaussian beams are not physical entities; that is, they do not exist in reality—much like plane monochromatic waves. However, laser sources and single-mode fibres can generate high-quality light beams that closely approximate this idealization. The so-called $\mathbf{M}^2$ analysis is commonly used to assess the quality of a laser by comparing it to an ideal Gaussian beam. The beam quality factor is defined as:

\begin{equation}
M^2 = \frac{W_0 \theta_0}{\lambda / \pi} = \frac{\pi W_0^2}{\lambda z_R},
\label{eq:m3}
\end{equation}
with $M^2 = 1$ for Gaussian beams and $M^2 > 1$ for non-ideal beams. For instance, gas lasers (e.g. Helium–Neon) often reach values of $M^2 < 1.1$; Ion lasers usually work in the range of $M^2 < 1.1–1.3$; collimated $TEM_{00}$ diode-laser beams present $M^2 \approx 1.1–1.7$; moving to high-energy multi-mode lasers hold $M^2$ factors as high as 3 or 4.

The $M^2$ factor determines the value of several experimental parameters. The spot size after a microscope's objective is given by:

\begin{equation}
    2W_{\text{\tiny{SS}}} = \frac{4M^2\lambda f}{\pi D},
    \label{eq:spotsizeobjective}
\end{equation}
where D is $\lambda$ is the excitation wavelength, $f$ is the focal length and $D$ is the diameter of the incoming beam. Alternatively, the confocal volume is defined as:

\begin{equation}
    V_{\text{\tiny{CONF}}} =  \sqrt\frac{2^5}{\pi^3} \frac{(M^2)^3 \lambda^3 f^4}{D^4}.
    \label{eq:confocalvolume}
\end{equation}

\subsubsection{Razor blade technique}

In order to study the beam profile the razor blade technique can be used. This approach allows for measurements of the intensity profile at multiple planes along the propagation axis without the need of specialized hardware. For this technique, the collimated beam of interest is focused onto a detector. The photodetection area should be large compared to the focused beam. 

\begin{figure}[h]
    \centering
    \includegraphics[width=\linewidth]{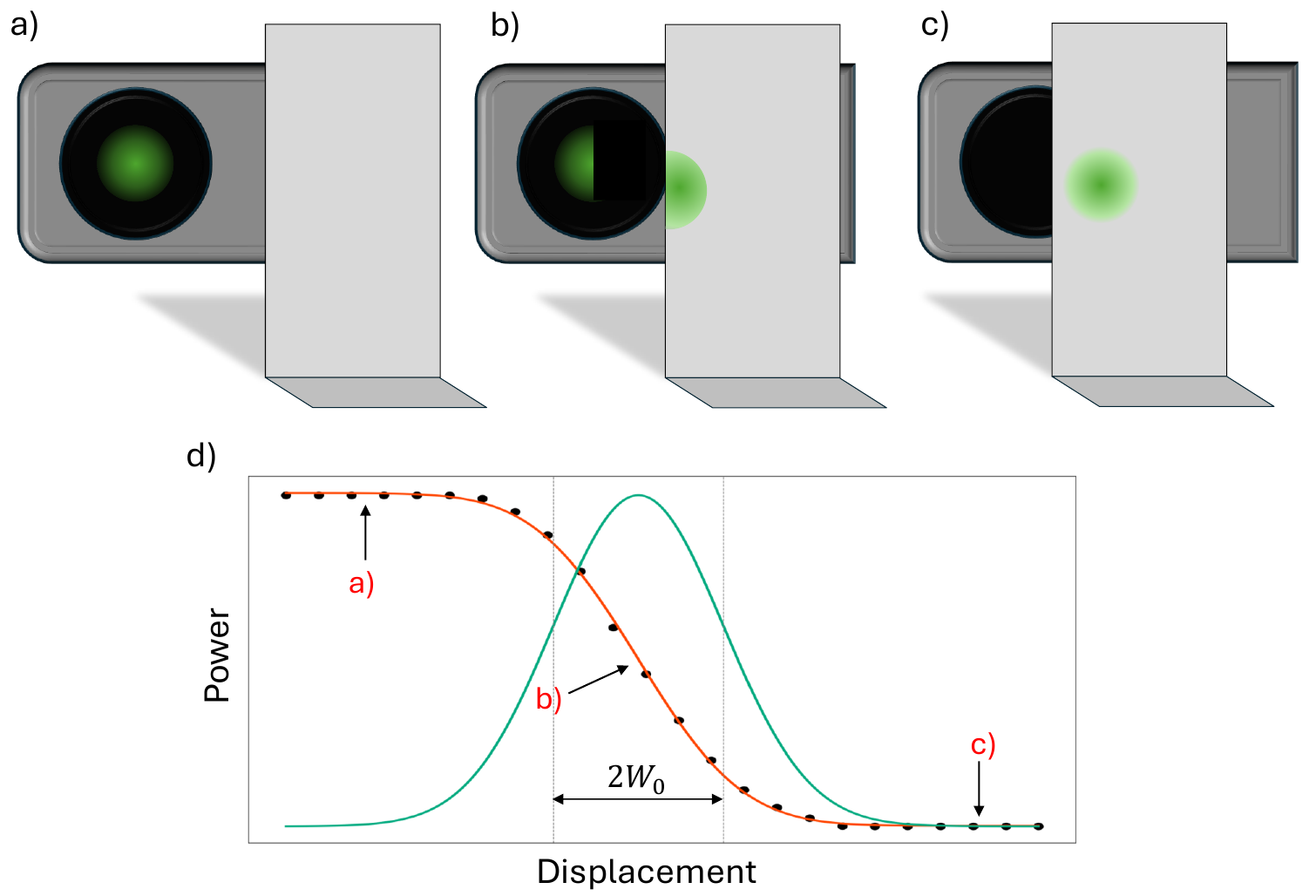}
    \caption{(a-c) Razor Blade technique to determine the beam width. A sharp barrier is slowly displaced to cover the studied beam that is hitting a powermeter. d) The power distribution will fall to zero following a complementary error function (orange curve) from which $W(z)$ is extracted. By taking the derivative of this function we recover the Gaussian intensity distribution (green curve).}
    \label{fig:razor2}
\end{figure}

For the measurements, a razor blade is be placed on top of a translation stage, enabling displacements perpendicular to the propagation axis and progressively blocking the beam (Fig. \ref{fig:razor2}). The transmitted power can be calculated by integration of Eq. \ref{eq:int}:

\begin{equation}
\begin{gathered}
P (x) = \int_{-\infty} ^\infty \int_x ^\infty I(x,y,z) dx dy = \\
= \frac{P_0}{2} \left[ 1 - \text{erf} \left( \frac{x \sqrt 2}{ W(z)} \right) \right] = \frac{P_0}{2} \cdot \text{erfc} \left( \frac{x \sqrt 2}{ W(z)} \right),
\label{eq:error}
\end{gathered}
\end{equation}
where erf is the error function and erfc is the complementary error function. The data is then fitted and the beam width $W(z)$ is extracted. By taking the derivative, the Gaussian intensity profile is recovered. Using the complementary error function to estimate $W(z)$ is preferable over the direct measurement of the Gaussian intensity profile. This is due to the finite size of detectors, which will affect the shape of the profile at the borders of the beam. This procedure must be repeated for several distances along the z axis. A higher sampling of the area around the focal spot helps the estimation of W$_0$, whereas for higher distances the linear behaviour can be successfully described with less points.

\subsection{Single-photon source}

Under non-ionizing excitation, optical pumping of NV centres drives transitions between the ground and the excited states. Since these transitions take a finite amount of time, the photons from the NV centre are emitted one at a time and hence it behaves as a single-photon source \cite{Beveratos2002_1, Beveratos2002_2}. This can be derived from the normalised second-order autocorrelation function, which is defined as \cite{Beveratos2002_1}:

\begin{equation}
    g^{(2)}(\tau) = \frac{\langle I(t)I(t+\tau)\rangle}{\langle I(t)\rangle^2},
\label{eq:g2}    
\end{equation}
where $g^{(2)}(\tau)$ quantifies the correlation between events separated by a time $\tau$. For photon statistics it will describe the probability of measuring a photon a time $\tau$ after having measured another one. Here a value of one implies they are completely uncorrelated. It is of special interest to study the value of $g^{(2)}(0)$, which will describe the possible correlation between two events occurring at the same time. For a single-photon source there is no probability of two events taking place at the same time, and so $g^{(2)}(0) = 0$. This is the anti-bunching effect, where photons are more evenly spaced than coherent light sources. The second-order autocorrelation function is related to the number of emitters as \cite{Beveratos2001, Beveratos2002_1}: 

\begin{equation}
    g^{(2)}(0,n) = 1 - \frac{1}{n}.
\label{eq:g2_0}
\end{equation}

Experimental noise can affect this measurement, but the single-emitter character can be assumed as long as $g^{(2)}(0) < 0.5$. For $\tau>0$, there can be trapping on the metastable state that leads to bunching effects where photons are no longer evenly spaced. This can be evidenced by values greater than unity for $g^{(2)}(\tau)$ \cite{Beveratos2002_1}.

\subsection{Polarization}

\begin{figure}[ht]
    \centering
    \includegraphics[width=\linewidth]{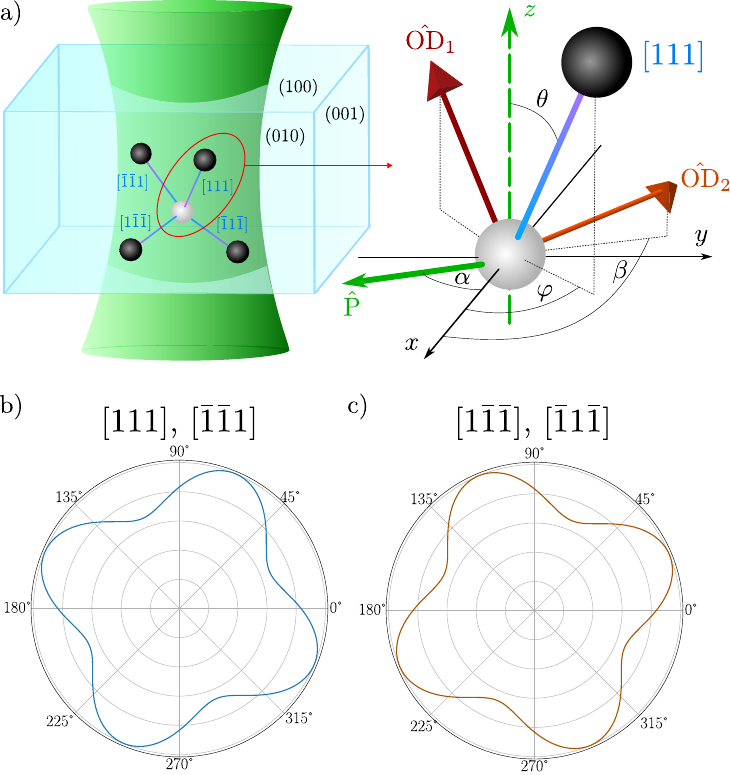}
    \caption{a) Schematic of all four crystallographic directions of NV centres inside a (100) cut diamond illuminated by green laser. A zoom of the [111] direction on the right shows the optical dipoles $\hat{\text{OD}}_1$ and $\hat{\text{OD}}_2$, perpendicular to the [111] axis and between each other. The laser irradiation is depicted as the x axis and the linear polarization as $\hat{\text{P}}$ lying on the xy plane. b) PL response of the [111] and [$\bar 1 \bar 1$1] orientations as a function of waveplate polarization $\alpha'$. c) Same signal for the [1$\bar 1 \bar 1$] and [$\bar 1 $1$\bar 1$] orientations. It appears shifted by 45º.}
    \label{fig:polarizacion-directions}
\end{figure}

Laser-equipped confocal microscopes deliver polarized light to the imaged sample which plays a crucial role in enhancing interaction specificity and photon counts when probing single NV centres. In a single-crystal diamond, NV centres can be oriented along one of four equivalent crystallographic [111] orientations and have been shown \cite{Chouaieb2019} to present two orthogonal optical dipole moments, perpendicular to their main axis. In practice, this implies that changing the polarization of light incident on a diamond will affect its emission characteristics. Thus, the photoluminescence (PL) level as a function of incident polarization will depend on the crystallographic orientation of the NV centre as well as the direction of laser irradiation. For a typical (100)-polished diamond and with optical irradiation normal to its surface (Fig. \ref{fig:polarizacion-directions}), the PL can be described as:

\begin{equation}
    I(\alpha, \theta, \varphi) \propto 1 -  \sin^2 (\theta) \cos^2(\varphi - \alpha),
    \label{eq:polarization}
\end{equation} 

\noindent where $\alpha$ defines the laser polarization angle in the xy plane, $\theta$ and $\varphi$ are the polar and azimuthal angles of the NV's main axis with respect to the z axis and $\beta$ is the azimuthal angle of one of the optical dipoles. For the remainder of the text, we will consider a polarization $\alpha' = \alpha/2$ to account for the rotation of light with a $\lambda$/2 waveplate, where a shift of angle $\alpha'$ results in a rotation of the polarization by $2\alpha'$.

For the case of a (100) cut diamond, the particular orientation $\beta$ of the optical dipoles does not have an effect on the NV's PL as long as the laser polarization sits on the xy plane. The symmetry of the four crystallographic orientations with the vertical axis gives rise to only two possible PL responses, that appear rotated by 45º with respect to each other. If the illumination comes from a different direction, like in the case of normal illumination on a (111) cut diamond, this degeneracy disappears and all four orientations can be distinguished optically.  

The difference in emission is modulated by the tilt angle $\theta$ whereas $\varphi$ determines the rotation of the signal. The difference will be maximum for light perpendicular to the NV's main axis, whereas light parallel to this axis will not exhibit polarization dependent absorption. For this reason, characterization of the PL's polarization dependence is required before any experiment is performed, since it can lead to weakening or even suppression of the received signal. Maximum emission is always preferred for sensing experiments with single NV centres in bulk diamond.

\subsection{Optical saturation}

Another major consideration point when preparing a confocal microscope optimized to address NV centres is the intensity of the excitation light. The fluorescence dynamics of the NV centres is determined by the rates defining the different decay routes. The NV centre energy scheme comprises spin triplets $^3A$ and $^3E$ and a pair of metastable intermediate singlet states. Spin-preserving transitions between the triplet states occur with green laser excitation, decaying with a wide spectrum in the red to infrared (600 nm to 850 nm), with a distinct zero phonon line at 637 nm. Transitions through the intermediate singlet states are weak and usually not detected optically but shelve the NV in a nonfluorescing state for approximately 300 ns and lead to a preferential population of the $m_S=0$ substate, which is exploited for spin readout \cite{Doherty2013}.

When the laser intensity is sufficiently high, the excitation rates can surpass the relaxation rates back to the ground state, leading to population trapping. This puts a limit on the emission capabilities, reaching a steady state where no more fluorescence can be extracted from the system.

This saturation behaviour can be described by the model:

\begin{equation}
    I(P) = a \cdot \frac{P}{P_{sat} + P} + b \cdot P,
    \label{eq:saturation}
\end{equation}
where $a$ accounts for the detection capabilities of the microscope as well as the luminescence yield of the colour centre \cite{Chapman2011}, $P_{sat}$ is the saturation power and the $b \cdot P$ term models background noise linearly. $P_{sat}$ will depend on the optical elements employed. High numerical aperture objectives can induce saturation powers lower than 100 $\mu$W \cite{Li2015}, whereas lower NAs can make $P_{sat}$ go above 1 mW \cite{Kurtsiefer2000}.

Although this model does not account for high-power processes such as two-photon absorption \cite{Khramtsov2017} or charge conversion \cite{Yan2013}, it does describe successfully a window of operation where these effects do not appear. Working at $P_{sat}$ will mostly ensure correct initialization and readout at moderate laser powers without ionization effects. 

However, the actual performance may vary across different systems, as factors such as spin noise, charge distribution, and surface damage can influence this operational window. 

\subsection{Spectrum}

At room temperature, NV centres show a broad emission spectrum with a phonon sideband extended from 650 up to 850 nm \cite{Clark1971}. There is a distinct Zero Phonon Line (ZPL) at 637 nm, although only around 4\% of the emission is concentrated at this wavelength \cite{Aharonovich2011}. The fluorescence lifetime in bulk diamond is around  10 ns for bulk diamond \cite{Kurtsiefer2000}, with higher values being found for other hosts such as nanodiamonds  \cite{Tisler2009}.

Even for a given colour centre, switching from the negatively charged to the neutral charged state is possible. This photo-induced phenomenon can alter the emission spectrum as NV$^0$ defects have their ZPL at 575 nm with their phonon sideband extending up to 650 nm \cite{Aharonovich2011}. A sample with both populations will have a broader spectrum from which an estimation of populations of each charged state is possible. For the case of photo-induced charge conversion, relaxation back onto the negatively charged state occurs in the dark with a time constant between 0.3 and 3.6 $\mu$s \cite{Gaebel2006}.

\subsection{Light intensity modulation}

Fast, nanosecond-scale intensity modulation of the green excitation laser is critical in experiments with NV centres, as it enables precise control over the preparation and readout of the NV spin state. This is most often accomplished by using a free-space acousto-optical modulator (AOM) operated externally to the laser. However, other solutions such as fibre-coupled AOMs or pulsed lasers may be used as well. 

AOMs are used to produce light pulses with nanosecond-scale rise times by exciting an optical medium via radio-frequency (RF) soundwaves to induce Bragg diffraction. These waves produce periodic changes in the material's refractive index, leading to a stratified medium in the acoustic wave propagation direction. For an incoming laser beam, the reflectance produced by a single layer of finite length L in the crystal is given by \cite{wiley}:

\begin{equation}
r_{\pm} = \pm i r_o \sinc \left[ (2k \sin \theta \mp q) \frac{L}{2 \pi} \right] e^{i \Omega t},
\label{eq:r}
\end{equation}
where $r_0$ is the maximum reflectance, $q$ and $k$ are the wavenumbers of the RF signal and the laser respectively, $\theta$ is the angle between the incident light and the layer, and the exponential term reads the phase acquired by the laser at different points of the layer. The $\pm$ solutions take in consideration that sound can be propagated in both directions.

\begin{figure}[h]
\includegraphics[width=\linewidth]{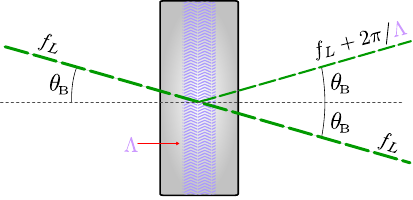}
\caption{AOM working principle. An acoustic wave of wavelength $\Lambda$ excites a crystal. Incoming beams with $\theta$ = $\theta_{\text{\tiny{B}}}$ will diffract, with the diffracted beam suffering an upshift in frequency equivalent to the acoustic frequency.}
\label{fig:braggv}
\end{figure}

The reflectance peaks for the Bragg angle $ \ 2 \Lambda  \sin \theta_{\mathcal{B}}= \lambda$, and rapidly falls to zero for a small deviance. The upshifted reflected wave leaves the medium whit a slightly increased frequency $E_r = r_{+} E \varpropto \exp (it(\omega + \Omega))$. This feature can be used in other set ups together with the laser source to obtain specific wavelengths.

The use of AOMs to produce pulses allows the delivery of steady amounts of power during its operation, resulting in a very convenient, stable excitation. This can be quantified through the Extinction Ratio (ER), which is defined as the ratio of the power output with the device switched on and off. However, the ability to generate such fast pulses typically comes at the expense of the deterioration of the Gaussian profile quality provided by the laser source. 

The performance of an AOM for laser pulsing will be determined by the incoming beam size. Complete diffraction of the beam requires the sound wave propagation all over the focused spot. For small beam sizes rise times will be shorter. In addition, the incoming RF signal must have a narrow bandwidth as well as have enough power to effectively produce diffraction.

Acousto Optic media can be found in different devices such as phase shifters, modulators or scanners. 

\section{Experimental setup}

\begin{figure*}[htbp]
  \centering
  \includegraphics[width=\textwidth]{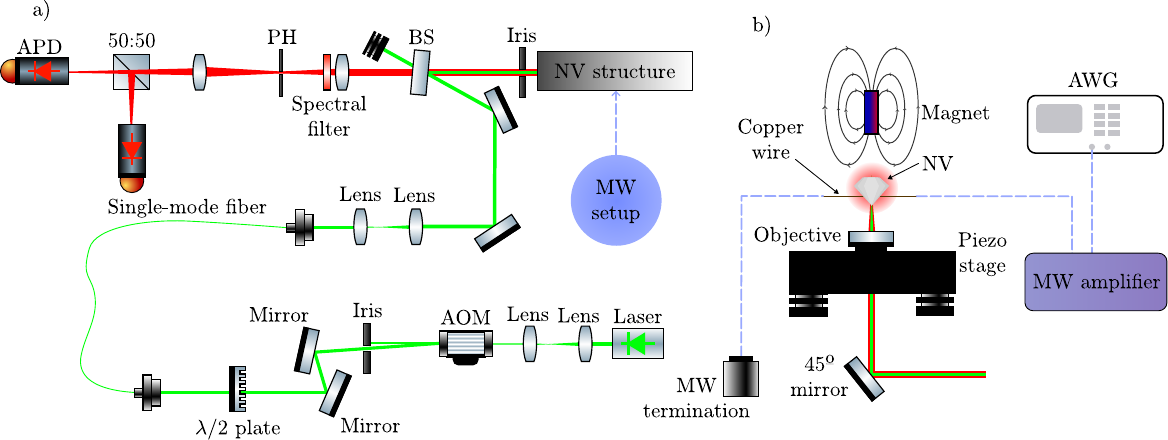}
  \caption{Experimental setup. a) Optical table configuration: a continuous laser is pulsed externally with an AOM and redirected towards a single-mode fibre. From there it is partially reflected on a beam sampler and sent to the diamond. The outcoming fluorescence is sent pass the beam sampler into a spectral filter to remove the laser contribution. Finally, it is focused onto an APD. A second detector is added for $g^{(2)}$ measurements. b) NV structure and MW setup: the laser beam is redirected upwards by a 45º mirror. A piezo stage controls the position of an oil objective to scan different areas of the diamond, affected by an external magnetic field. The MW signal is generated with an AWG and amplified before being irradiated by a copper wire next to the colour centres. The signal is then sent to a MW ground to avoid reflections.}
  \label{fig:imagen-ancha}
\end{figure*}

Confocal microscopes for quantum sensing experiments with NV centres typically have a form of custom-built optical setups on an optical table, with epifluorescence configuration, i.e., with illumination and fluorescence light collection occurring through the same microscope objective, as shown in Fig. \ref{fig:imagen-ancha}. 

In this configuration, a continuous laser serves as the light source, leveraging its temporal intensity stability for long-duration experiments. It is externally pulsed using AOM, enabling precise customization of pulse sequences with nanosecond resolution. Additionally, this arrangement offers a high extinction ratio, which is crucial for the proper re-initialization of NV centres and enables their coherent manipulation in the darkness, typically using microwave fields, between the optical interrogation pulses. Pulsed lasers can also be suitable for such experiments, but continuous lasers that are pulsed externally offer better customization of pulse sequences while maintaining a more stable output.

The Gaussian profile of the beam (M$^2$ = 1.1 $\pm$ 0.1 ) is severely affected after the AOM due to its diffractive nature, resulting in a superposition of transmitted modes.  A single-mode fibre is added to bring said feature back. Alignment on bare fibres can become a challenging task unless recursive methods are used. Two mirrors with kinematic mounts are placed before to redirect the beam into the fibre with an air objective. Additionally, the fibre is cleaved with specific equipment to increase transmittance and placed on a 5-axis mount to manoeuvrer the position of the tip and correct its tilting.
\begin{figure}[H]
    \centering
    \includegraphics[width=\linewidth]{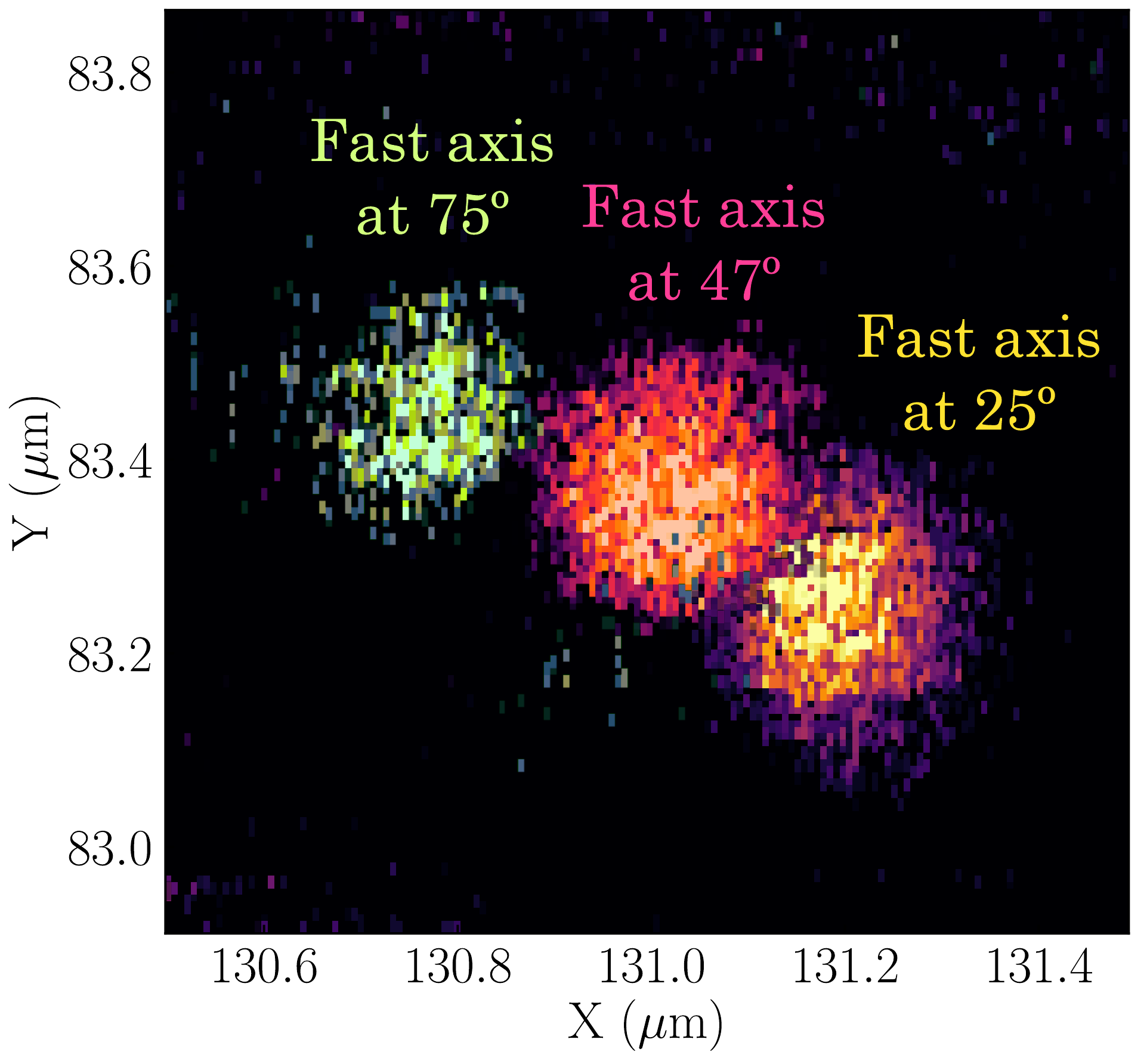}
    \caption{Superposition of 3 confocal scans of a single NV centre for different orientations of the $\lambda$/2 waveplate. The deviation induced by the waveplate results in misalignment of the setup and a shift of the produced image. Fluorescence is completely lost for values outside this region, making it crucial to have reference elements to redirect your beam when performing this type of measurement.}
    \label{fig:polarization-mal}
\end{figure}
A $\lambda/2$ plate will control the polarization of the naturally horizontal light (1:100) coming from the laser. It is placed in front of the fibre to use it as a fixed path to reroute light. Waveplates like the one we employ can induce deviations up to 20 arcmin when mounted. For a confocal microscope with high-power optics this means complete misalignment and loss of signal coming to the detector (Fig. \ref{fig:polarization-mal}).

Two telescopic lenses are placed after the lens, slightly reducing the beam size to underfill the entrance pupil diameter (8.52 mm for our 60X oil immersion objective). Over-filled pupils can lead to diffraction effects at the focal spot, which is avoided by underfilling the pupil at the cost of a slight loss of radial resolution \cite{Webb1996}. The laser is then rerouted towards a BSF20-B beam sampler, with anti-reflection (AR) coating in the 650-1000 nm range. Its reflectivity is polarization dependent with a maximal value for the AR coating range at 45º of incidence (0.5 $\%$) and minimal for near vertical incidence. The mirrors position the beam at around 5º with respect to the beam sampler to reduce this effect. Although much of the input power is transmitted through the beam sampler and lost, it is a convenient choice because fluorescence from the NV centres is also transmitted, ensuring maximum recollection.

The beam is redirected upwards in an inverted microscope configuration. In our setup, scanning is achieved by moving the objective, which is controlled by a piezoelectric stage. These devices, which can be used to displace either the objective or the sample itself, can work with sub-nanometer resolution. Their high stability is crucial to ensure that the microscope works at the optical resolution limit while scanning a sample.

MW irradiation is usually achieved via striplines designed on the Printed Circuit Board (PCB) where the diamond is placed. Their design needs to be compatible with the type of illumination of the microscope. For an inverted microscope, highly annealed copper wires can also be used. The wire must to be thin enough to fit inside the objective's working distance, and a MW termination needs to be placed to avoid reflections of the signal back to the instruments. The MW pulses are sent from an Arbitrary Waveform Generator (AWG) that creates the sequences and coordinates multiple instruments. They later pass an amplifier that increases the signal amplitude. For reference, a 15 to 20 dBm MW signal irradiated from a thin copper wire on a single NV separated 40 $\mu$m can produce oscillations from 40 to 100 ns of period.

Following excitation, a portion of the emitted fluorescence is collected by the objective and directed through the beam sampler. A longpass filter is then applied to eliminate any residual green light, and the remaining fluorescence is focused by an achromatic lens into a pinhole, ensuring a minimal focal spot size across the entire NV spectrum bandwidth. To maximize fluorescence collection, a red-to-infrared anti-reflection (AR) coating is used. Finally, the light is directed to an avalanche photodiode for photon counting.

\section{optimization of laser parameters}

We will implement a protocol composed of several experimental measurements in order to optimize the laser irradiation. First, the gaussianity of the beam will be measured using the razorblade technique before the oil objective to quantify the loss of the ideal profile after passing several optical elements. After that, a single NV centre is selected and its second-order correlation function is obtained. Next, the saturation of the NV centre will be obtained by measuring its fluorescence as a function of laser power and fitting a model where background noise is subtracted to find the optimal excitation. The polarization of the beam will be characterized by rotating it with a $\lambda/2$ waveplate and measuring the fluorescence response of NV centres. The maxima of emission will be localized with a fit to the theoretical model. A spectra will be obtained to characterise the emission. Finally, the power of the RF pulses will be varied through the AOM's driver device and the performance of the pulses will be assessed by studying the rise times, fall times, extinction ratios and contrast of ODMR signals.

\subsection{Gaussian intensity mode}

\begin{figure}[ht]
    \centering
    \includegraphics[width=\linewidth]{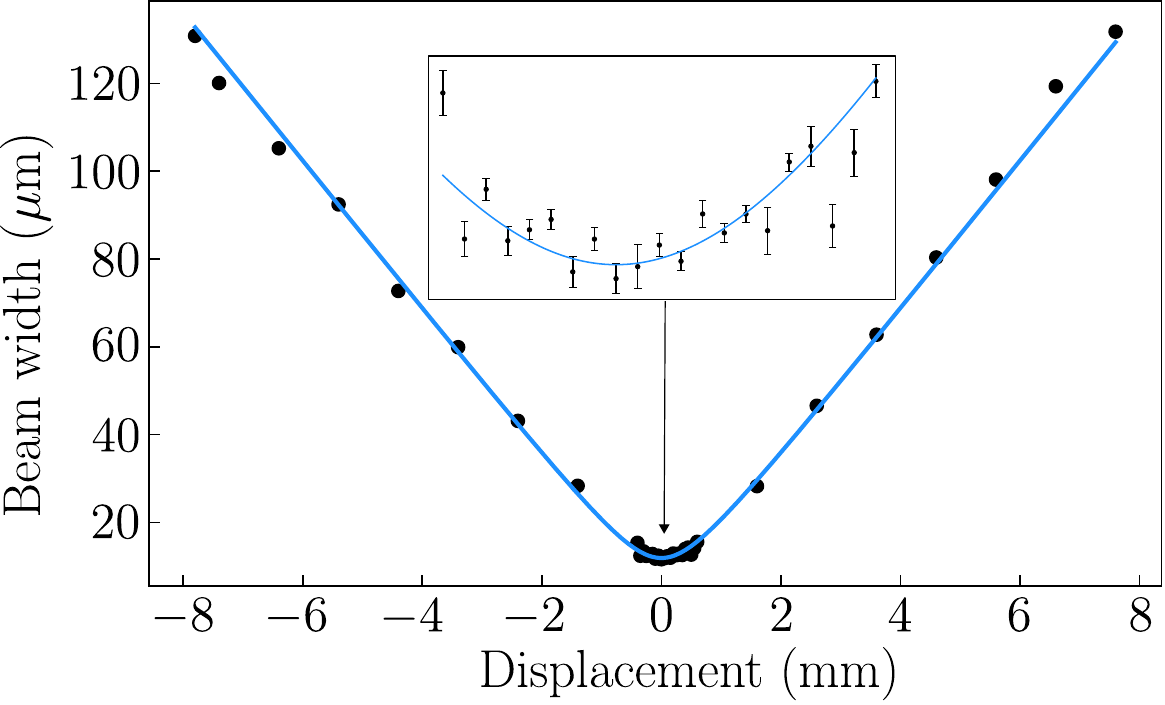}
    \caption{Beam width $W(z)$ as function of distance z, fitted to Eq. \ref{eq:beam} (blue line). We find a minimal beam width of 11.9 $\pm$ 0.5 $\mu$m with a Rayleigh distance $z_R$ = 700 $\pm$ 30 $\mu$m.}
    \label{fig:beamw}
\end{figure}

After measuring the complementary error function with the razorblade method and deriving the beam width for each value of displacement on the z axis, we show the results in Fig. \ref{fig:beamw}. 

The experimental data closely follows the theoretical model, exhibiting linear behaviour both before and after the focal plane. From the fitting of Equation \ref{eq:beam} we find $W_0 = 11.9 \pm 0.5 \text{ $\mu$m}$ and $z_R = 700 \pm 30 \text{ $\mu$m}$. From Eq. \ref{eq:m3} we obtain $M^2 = 1.19 \pm 0.15$. 

Following Eqs. \ref{eq:spotsizeobjective} and \ref{eq:confocalvolume} for our microscope parameters, we find $2W_{\text{\tiny{SS}}} = 300 \pm 40$ nm and $V_{\text{\tiny{CONF}}} = (5\pm 2)\cdot10^{-3}$ $\mu$m$^3$.

\subsection{Second-order autocorrelation function}

In order to prove that we are working with a single NV centre, a value lower than 0.5 is needed for the normalized second-order autocorrelation function at $\tau$=0. For this measurement, we will sample the timing between events from -150 to 150 ns approximately in intervals of 400 ps.

\begin{figure}[h]
    \centering
    \includegraphics[width=\linewidth]{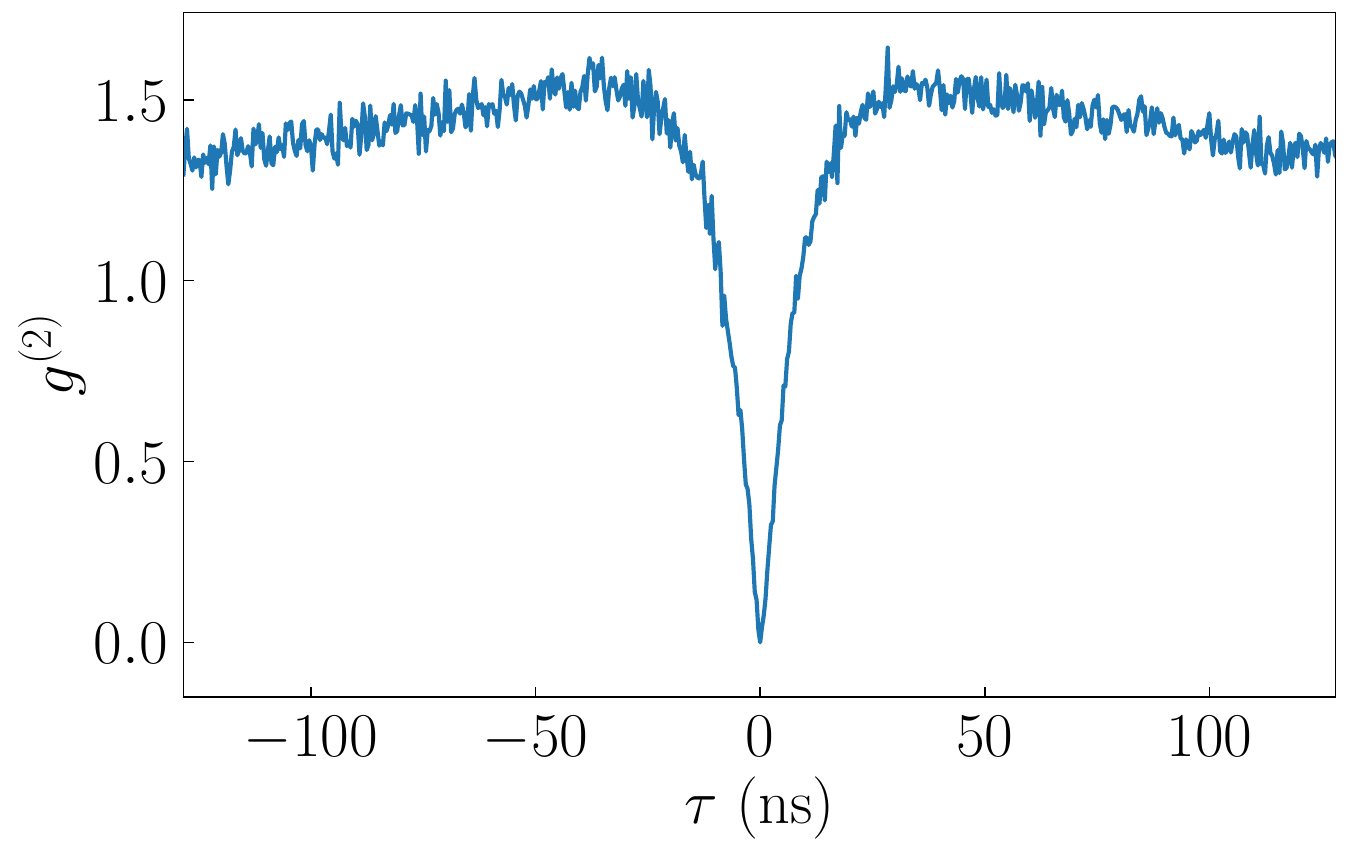}
    \caption{Second-order autocorrelation function $g^{(2)}$ for a single NV centre. A dip evidences the colour centre is behaving as a single photon source.}
    \label{fig:g2}
\end{figure}

A clear anti-bunching effect can be seen from a distinct dip at $\tau = 0$ ns, where a value of 10$^{-17}$ for $g^{(2)}$ confirms the single photon source character of the selected NV centre.

\subsection{Saturation}

\begin{figure}[ht]
    \centering
    \includegraphics[width=\linewidth]{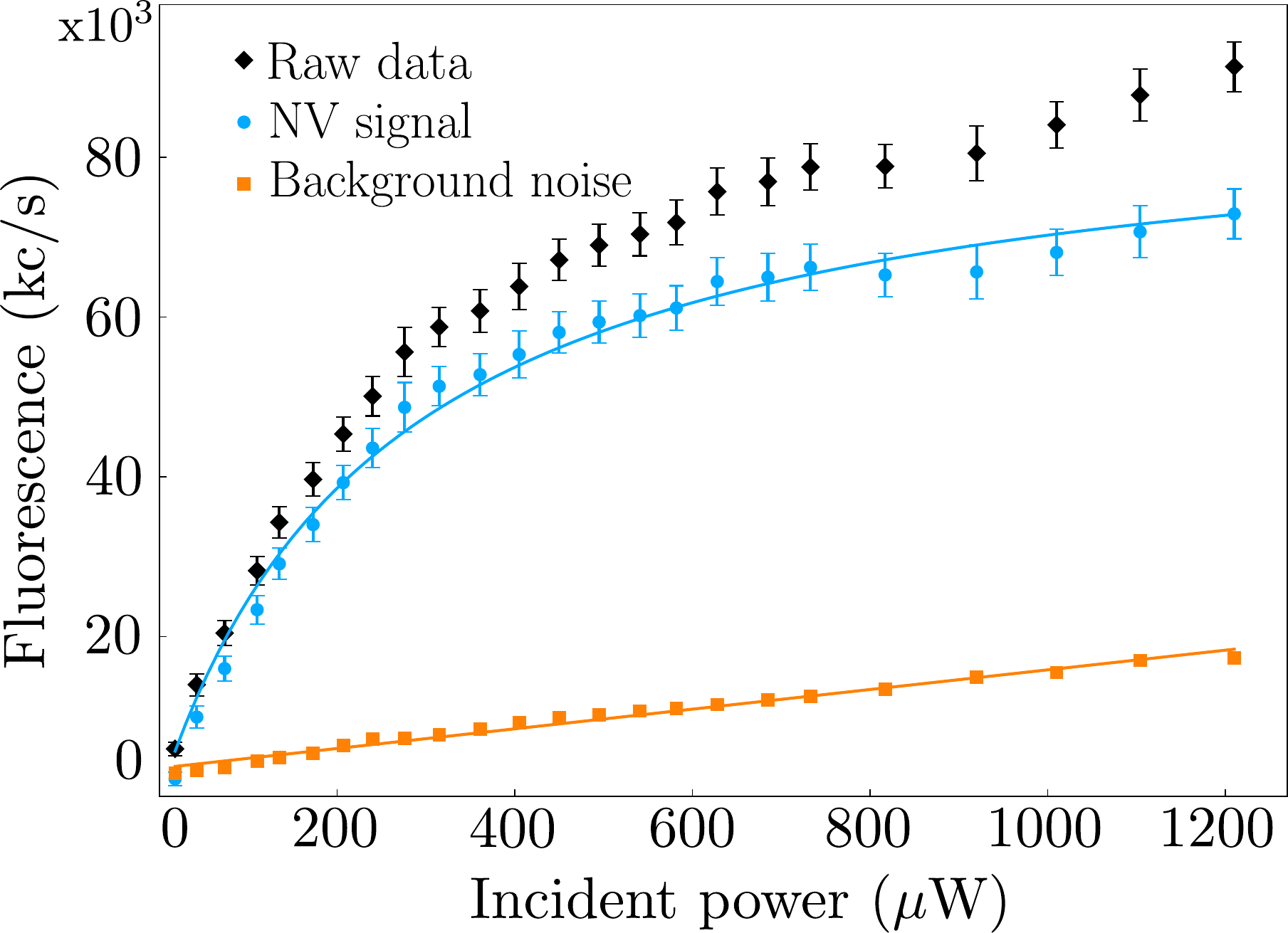}
    \caption{Saturation measurement of a single NV (black diamonds) where background noise (orange squares) is modelled linearly and subtracted to obtain the response of the NV centres (blue points). We find a saturation power of 258 $\pm$ 16 $\mu$W.}
    \label{fig:saturation}
\end{figure}

To measure the saturation of a single NV centre, its photoluminescence properties are studied under different laser powers. The photocounts collected by the APD are averaged for 10 to 20 seconds at each value of incident laser power. The background noise is modelled linearly and subtracted at each point. The results are shown in Fig. \ref{fig:saturation}.

After fitting of the model described by Eq. \ref{eq:saturation} we find a saturation power $P_{sat} = 258 \pm 16 \text{ $\mu$W}$. For a population with the same implantation characteristics, saturation values will lie closely to that of the NV measured. Thus, a single saturation measurement can serve as a solid reference for all emitters in a sample.

\subsection{Polarization}

\begin{figure}[h]
    \centering
    \includegraphics[width=\linewidth]{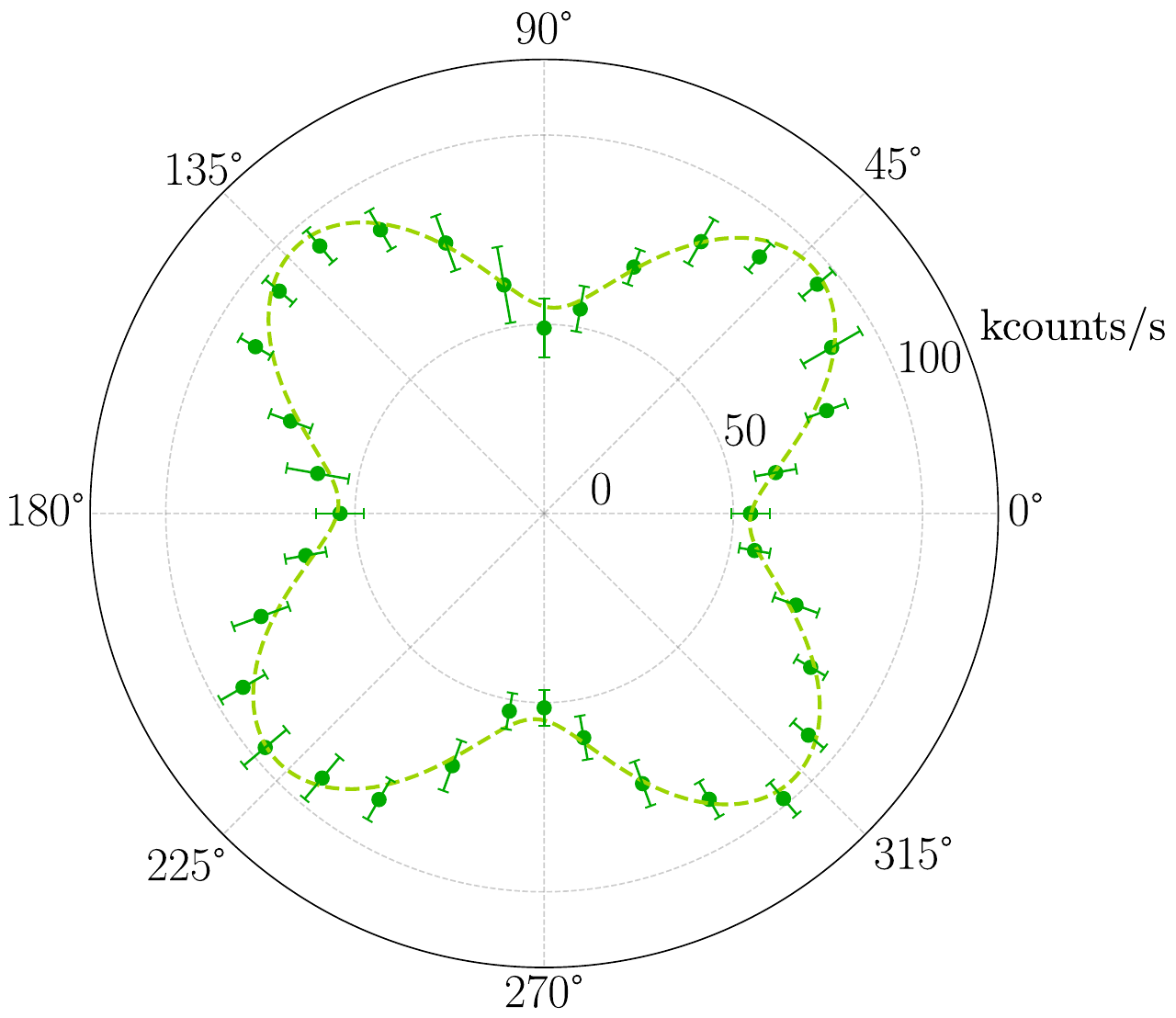}
    \caption{Fluorescence response of an NV centre to the change in polarization induced by the rotation of a $\lambda/2$ waveplate. The waveplate is rotated in steps of 10º and fluorescence is measured. Since a rotation of the waveplate of angle $\theta$ creates a dephasing of 2$\theta$ in the incident polarization, 4 maxima appear. A fit of Eq. \ref{eq:polarization} is performed (green dashed line) to find the maxima of emission.}
    \label{fig:polarization-nv}
\end{figure}

To measure the polarization response of a single NV, its photoluminescence is again sampled for different polarization angles of the incident laser beam. One $\lambda$/2 waveplate is rotated in intervals of 10º and photocounts are detected. After measuring the fluorescence from a single NV centre and averaging for 20 seconds for each point we response shown in Fig. \ref{fig:polarization-nv}

The output signal does respond accordingly to the theoretical model of Eq. \ref{eq:polarization}. The maxima are located at multiples of 45º, with the fitting returning an exact value of $41.2 \pm 0.6$º.

Although this value will change for other NV centres in the same population in different crystallographic directions, it will maximize the fluorescence response of those with this specific orientation. Those colour centres not aligned will still be visible and the polarization can be readjusted to maximize their response as well, which will lie between 0 and 45 degrees.

\subsection{Spectrum}

To confirm and characterize the optical emission of NV centres, we measure its fluorescence spectrum under green laser excitation at a power below the saturation threshold $P_{\text{\small{sat}}}$. This ensures minimal photo-induced charge conversion \cite{Yan2013} and avoids spectral distortion due to high-power effects. 

\begin{figure}[htb!]
    \centering
    \includegraphics[width=\linewidth]{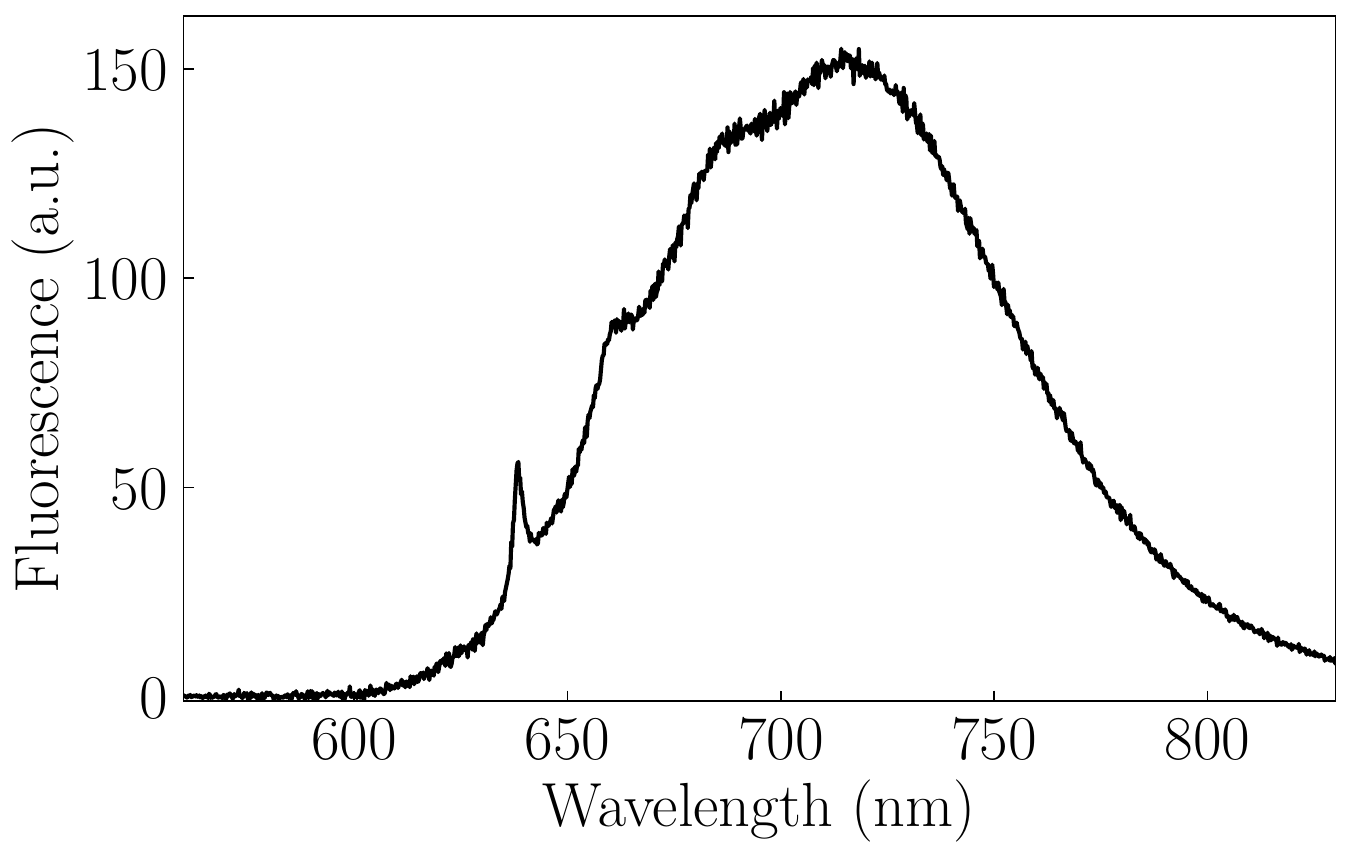}
    \caption{Fluorescence emission spectrum of a single NV centre. The ZPL transition at 637 nm is visible with the maximum of emission around 715 nm.}
    \label{fig:spectrum}
\end{figure}

The resulting spectrum, shown in Fig. \ref{fig:spectrum}, reveals a broad phonon sideband extending from approximately 600 nm to 850 nm, with a distinct Zero Phonon Line (ZPL) at 637–638 nm. This ZPL accounts for only a few per-cent of the total emission, yet serves as a key spectral fingerprint of the negatively charged NV$^-$ state. Importantly, no significant fluorescence is observed in the 550–600 nm range, indicating the absence of NV$^0$ emission. This confirms that the NV centre remains in the desired, negative charge state throughout the measurement.  Moreover, the absence of NV$^0$ fluorescence and light from other background sources is critical for ensuring stable spin readout and high-contrast ODMR signals. 

\subsection{Laser pulses}

For this last step of the protocol, we study the quality of laser pulses by varying the RF power in a range in accordance with the damage limitations advised by the fabricant. For our device, this implies a range between -16.4 and -19.3 dBm. The pulse is evaluated by measuring the power transmittance, the extinction ratio, the rise and fall times and the contrast from an ODMR experiment with the same emitter under the same pulse extraction configuration. Here, the contrast is computed as:

\begin{equation*}
    \mathcal{C_{\text{\tiny{ODMR}}}} = \frac{I_{\text{\tiny{AVG}}} - I_{\text{\tiny{MIN}}}}{I_{\text{\tiny{AVG}}}},
\end{equation*}
where $I_{\text{\tiny{AVG}}}$ represents the average of the first and last five points of the sequence and $I_{\text{\tiny{MIN}}}$ represents the minimum of the function. This method represents a direct calculation of contrast in our measurement and is preferable when the signal is averaged less times and the Lorentzian fits are not as reliable. 

\begin{figure}[h]
    \centering
    \includegraphics[width=\linewidth]{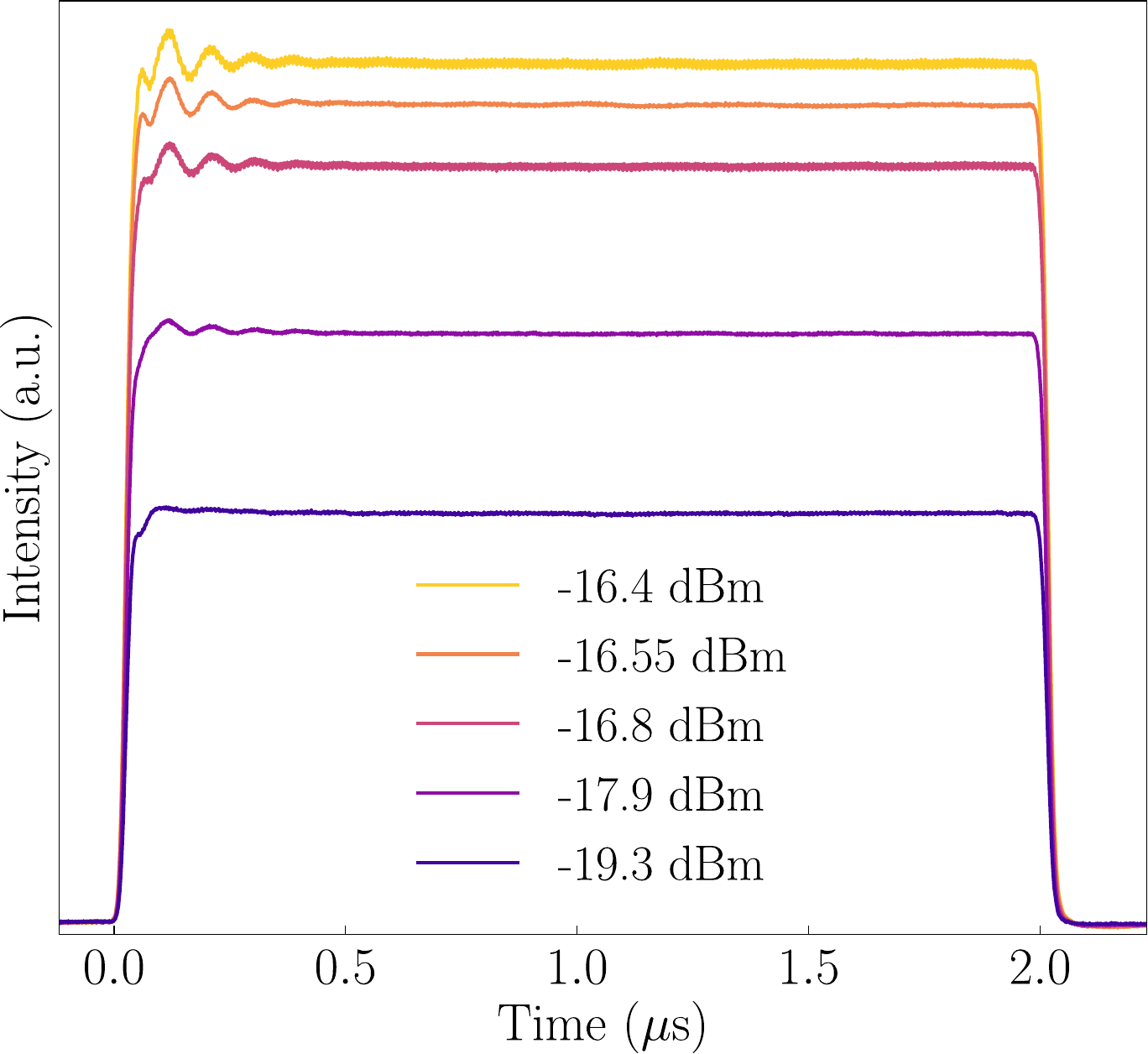}
    \caption{2 $\mu$s laser pulses obtained from different values of power excitation at the AOM. The higher values reproduce small ripples at the beginning of the pulse, coming from the triggering of the aom driver.}
    \label{fig:aom}
\end{figure}

Fig. \ref{fig:aom} shows digital pulses of 2 $\mu$s of duration for different RF powers. The top signal is vertically displaced for better visibility, otherwise laying on top of the second measurement. Interestingly, some ripples appear for higher powers, originated from the digital source that triggers the AOM. They diminish as less RF power is delivered.

\begin{table}[h]
    \centering
    \begin{tabular}{
        c@{\hskip 0.3cm}
        c@{\hskip 0.3cm}
        c@{\hskip 0.3cm}}
        \toprule
        \textbf{RF power(dBm)} & \textbf{Power transm.(\%)} & \textbf{ER} \\
        \midrule
        -16.4 & 45.5 & 460760 \\
        -16.55 & 43.25 & 487323 \\
        -16.8 & 35.87 & 478333 \\
        -17.9 & 25.62 & 394231 \\
        -19.3 & 11.62 & 206667 \\
        \multicolumn{3}{c}{} \\[-1.8ex]  
        \toprule
        \textbf{Rise time(ns)} & \textbf{Fall time(ns)} & \textbf{Contrast(\%)} \\
        \midrule
        28.8(0.3) & 29.6(0.3) & 19.8 \\
        29.6(0.3) & 28.8(0.3) & 20.4 \\
        30.4(0.3) & 28.0(0.3) & 20.3 \\
        29.6(0.3) & 28.0(0.3) & 19.3 \\
        27.2(0.3) & 28.0(0.3) & 19.6 \\
        \bottomrule
    \end{tabular}
    \caption{Power transmittance, extinction ratio (ER), rise and fall times and contrast as a function of RF power. The values chosen for this parameter are set to the range of operation recommended by the fabricant.}
    \label{tabla_att}
\end{table}

Analysing the results of Table \ref{tabla_att}, power transmittance increases with the amplitude of the RF signal. Nevertheless, the extinction ratio does not follow this clear behaviour, with the maximum being found for the second highest power. Similarly, rise and fall times are not uniform, with the best performance being found for the lowest excitation power. This might be due to the oscillations present at high RF amplitude. Finally, we see very similar values of contrast for our ODMR signal, indicating that these effects do not seem to affect the signal quality.

In conclusion, low RF powers produce better responsivity without external oscillatory noise at the cost of a worsened ER. From the point of view of signal contrast, none of these effects seem to have a noticeable impact.

\section{Results}

The protocol that we have applied to our microscope can now be tested with some of the functionalities that compose regular practice for quantum sensing with NV centres. First and foremost, XY scans of single NVs must show good enough resolution and SNR for quick identification and correct initialization. This becomes critical when working with samples where we have different refractive indices and more fluorescence is lost on the path back. 

\begin{figure*}[htbp!]
  \centering
  \includegraphics[width=\textwidth]{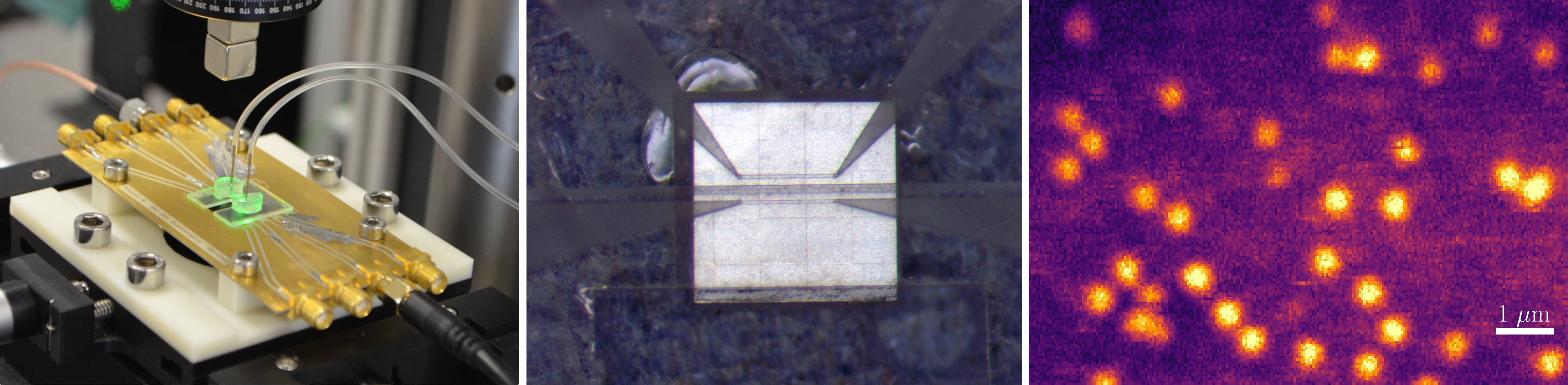}
  \caption{Confocal imaging of single NVs in microfluidic device. On the left, a silica glass chip for microfluidic experiments mounted on top of a gold PCB. The chip is illuminated with green laser. Two needles installed on the inlet and outlet entries lead a flow of the studied fluid through the tubes into the chip. On the centre, a microscope image of the 2x2 mm CVD-grown diamond installed on top of the microfluidic channels. On the right, an XY scan of single NV centres at a concentration of $10^9 \frac{^{15}\text{N}^+}{\text{cm}^2}$. The NVs appear distinguishable despite the changes of refractive index between surfaces.}
  \label{fig:microfluidics}
\end{figure*}

To test this issue we perform confocal imaging of a 2x2 mm bulk diamond introduced into a silica glass chip with etched microchannels for microfluidic experiments. This is a specific example of how confocal microscopy is employed for a given experiment, but other systems will follow the same principle. The setup and scan is shown in Fig. \ref{fig:microfluidics}. In this case, the immersion oil-glass-microfluid interfaces as well as the geometry of the channels complicate excitation and recollection of fluorescence. Nevertheless, NV centres still appear distinguishable as shiny spheres near the limit of resolution. A 2D-Gaussian filter is applied for blur correction appearing from dust deposited on the microfluid being sent.

\begin{figure}[h]
    \centering
    \includegraphics[width=\linewidth]{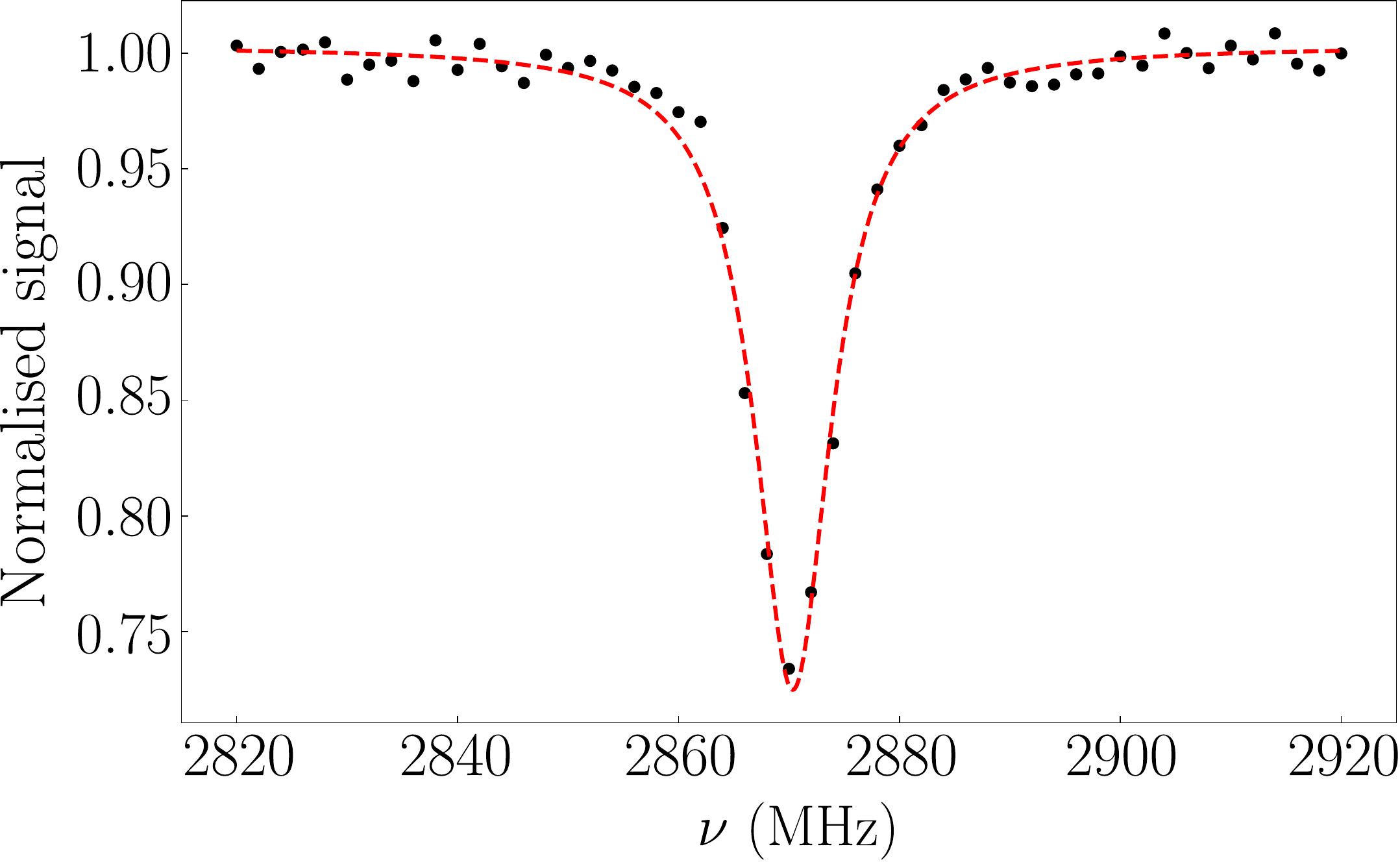}
    \caption{Optically Detected Magnetic Resonance (ODMR) experiment showing a contrast of 27.9 $\pm$ 0.6\%, near the 30\% limit given by electronic dynamics. The contrast is taken as the amplitude of the Lorentzian fit.}
    \label{fig:odmr}
\end{figure}

Fluorescence recollection and effective pulsing will also affect the quality of obtained signals. To test the capabilities in this area, we performed an ODMR experiment to determine the resonant transition frequency of our NV centre. The total contrast that can be extracted from the signal is limited by the NV centre's electronic dynamics, having a maximum of around 30\% \cite{Barry2020, Doherty2013}. This value is usually decreased by signal mixing and inefficient pulsing. Our measurements are shown in Fig. \ref{fig:odmr}. After repeating the protocol around 500.000 times, we find a contrast of 27.9\% with a well defined Lorentzian shape. This shows good quality of laser control and efficient pulsing. In this case, the sequence was repeated enough times for the amplitude of the Lorentzian fit to work as a reliable representation of the fluorescence contrast. 

In conclusion, we have been able to study the main parameters describing the intensity profile, photoluminescence properties and pulsed character of our excitation laser. We have given a theoretical frame for the Gaussian mode and how to measure it with the $M^2$ formalism and we have put it into practice with the razorblade technique in our setup. We measured the second-order autocorrelation function for a single emitter and we also optimized both the excitation intensity and the beam polarization before measuring the optical spectrum. In addition, we were able to study the shape of the pulses while varying the attenuation of the incident acoustic wave. The method that we have presented to study these parameters is both accessible without specialized instrumentation as well as efficient in time and efficacy, making it easily reproducible for any confocal microscope with these characteristics. 

\bibliographystyle{ieeetr}
\bibliography{gaussianarticle_mendeley.bib}
\end{document}